\documentclass[12pt]{iopart}
\usepackage[ruled,vlined]{algorithm2e}

\usepackage{iopams}  
\usepackage{xcolor}
\usepackage{graphicx}
\usepackage{subcaption}
\usepackage{bigdelim, array}
\usepackage{multicol}
\usepackage{multirow}
\usepackage[toc,page]{appendix}

\begin{document}

\title[]{Harnessing the Power of Gradient-Based Simulations for Multi-Objective Optimization in Particle Accelerators}

\author{Kishansingh Rajput$^{*, 1, 2}$, Malachi Schram$^{1,3}$, Auralee Edelen$^{4}$, Jonathan Colen$^{5, 6}$, Armen Kasparian$^{1}$, Ryan Roussel$^{4}$, Adam Carpenter$^1$, He Zhang$^1$, Jay Benesch$^1$}
\address{$^1$ Thomas Jefferson National Accelerator Facility, Newport News, VA 23606, USA}
\address{$^2$ Department of Computer Science, University of Houston, Houston, TX 77204, USA}
\address{$^3$ Department of Computer Science, Old Dominion University, Norfolk, VA 23529, USA}
\address{$^4$ SLAC National  Laboratory, Menlo Park, CA 94025, USA}
\address{$^5$ Joint Institute on Advanced Computing for Environmental Studies, Old Dominion University, Norfolk, VA 23539, USA}
\address{$^6$ Hampton Roads Biomedical Research Consortium, Portsmouth, Virginia, 23703 USA}
\ead{$^*$ kishan@jlab.org}
\vspace{10pt}
\begin{indented}
\item[]August 2024 
\end{indented}

\begin{abstract}
Particle accelerator operation requires simultaneous optimization of multiple objectives. Multi-Objective Optimization (MOO) is particularly challenging due to trade-offs between the objectives. 
Evolutionary algorithms, such as genetic algorithm (GA), have been leveraged for many optimization problems, however, they do not apply to complex control problems by design. 
This paper demonstrates the power of differentiability for solving MOO problems using a Deep Differentiable Reinforcement Learning (DDRL) algorithm in particle accelerators. We compare DDRL algorithm with Model Free Reinforcement Learning (MFRL), GA and Bayesian Optimization (BO) for simultaneous optimization of heat load and trip rates in the Continuous Electron Beam Accelerator Facility (CEBAF). 
The underlying problem enforces strict constraints on both individual states and actions as well as cumulative (global) constraint for energy requirements of the beam. 
A physics-based surrogate model based on real data is developed. 
This surrogate model is differentiable and allows back-propagation of gradients. 
The results are evaluated in the form of a Pareto-front for two objectives. 
We show that the DDRL outperforms MFRL, BO, and GA on high dimensional problems.
\end{abstract}

\section{Introduction}
\label{ch:introduction}

Particle accelerators are intricate, high-energy machines comprised of numerous specialized components. 
To ensure efficient operation and precise control, accelerator operators must meticulously adjust multiple component settings to fine-tune the machine.

Jefferson Laboratory's primary particle accelerator, Continuous Electron Beam Accelerator Facility (CEBAF) \cite{CEBAF}, is comprised of two anti-parallel superconducting radiofrequency (SRF) linear accelerators (linac) to accelerate electrons.  
Each linac contains 25 cryomodules comprising of 200 SRF cavities.
These cavities are kept at 2K temperature to maintain superconductivity.
The cryomodules are filled with liquid helium regulated by a Central Helium Liquefier (CHL).
These cavities are individually controlled and each has its own unique operating characteristics.  
An ever-present challenge faced by operations staff is determining the best way to distribute Radio Frequency (RF) gradients across these cavities to meet the required experiment energy gain and simultaneously minimize negative impacts to CEBAF and experiments.
Two aspects of SRF operations that are directly controlled through the gradient distribution is the heat load imposed upon the cryogenics system through RF operations and the number of Fast Shut Down (FSD) trips initiated by the RF system.
Optimizing the RF heat load and FSD trips is particularly important since it would lower the wear and tear in CHL and reduce machine downtime respectively.

This is a multi-objective optimization (MOO) problem, in which there is a tradeoff between the heat load and the number of FSD trips. 
In MOO, the full optimal set of tradeoffs is known as the Pareto-optimal front (or Pareto front). 
This front defines the best quality solution that can be achieved for a given objective without reducing the quality of a competing objective. 

The problem we are considering thus becomes a question of how to efficiently find the Pareto-optimal front defining the tradeoff between the FSD trips and RF heat load, ideally in a way that will also translate from offline system analysis to real-world facility operation.
In addition to RF heat load and FSD trips, the gradient distribution needs to produce an energy within a very small tolerance for proper steering of the beam.
This introduces a global hard constraint on the optimal solution sets, making the problem very challenging especially when considering the large number of cavities in a linac.
From an operational point of view, a quickly-converging algorithm on this high-dimensional MOO problem is desired.

In this paper, we use a surrogate model of the CEBAF RF heat load and trip rates based on historical data to train and compare MFRL based Conditional Multi-Objective Twin Delayed Deep Deterministic Policy Gradient (CMO-TD3)~\cite{TD3}, NSGA-II \cite{nsga2}, Multi-Objective Bayesian Optimization (MOBO)~\cite{roussel2021multiobjective}, and Conditional Multi-Objective Deep Differentiable Reinforcement learning (CMO-DDRL)~\cite{DDRL} algorithms in the offline multi-objective optimization task described above. 
We investigate the performance of these algorithms in terms of time- and sample- efficiency, as well as solution quality. 
We also assess how the performance changes with problem dimensionality (i.e. number of RF cavities used). 
Finally, embedded in this investigation, we quantitatively assess the impact of using a differentiable system model (as is done in DDRL), in contrast to algorithms that do not use such a model (NSGA-II, TD3, MOBO). 
We also highlight some advantages the DDRL approach has in terms of being able to actively scan the Pareto-optimal front in a control setting. 
This investigation highlights the importance of sample acquisition speed in determining appropriate algorithms to use in MOO. 
It also provides a useful case-study for the accelerator community, which will aid researchers in determining approaches to use for problems on other systems. 

The paper is organized as follows: Section~\ref{ch:previous} describes the latest work in AI-based optimization for particle accelerators, Section~\ref{ch:problem} describes the CEBAF optimization  challenge, and the methods and results are described in Section~\ref{ch:method} and Section~\ref{ch:results} respectively. 
We close the discussion with some insights and future outlook in Section~\ref{ch:conclusion}.
\section{Previous work}
\label{ch:previous}

Machine Learning (ML) has been instrumental in solving complex challenges in many science and engineering applications. 
ML techniques have been applied to a variety of particle accelerator applications including anomaly detection and prediction \cite{Rajput_2024, Blokland:2021onk, edelen_anomaly_2021, ALANAZI2023100484, 9658806}, 
and surrogate modeling \cite{Kafkes:2021jse, schram2023uncertainty}. 

In the context of determining Pareto-optimal tradeoffs between competing objectives (and particularly in an offline or design setting), Genetic Algorithm (GA) have historically been the algorithm of choice in the accelerator community.
In the past, GAs have been explored to assist in gradient distribution in CEBAF linacs \cite{CEBAF_GA} to simultaneously minimize both heat load and trip rates while maintaining required energy gain. 
Charged particle optics design (CPOpt), is a multi-objective genetic algorithm (MOGA)  framework using NSGA-II~\cite{deb2002fast} to optimize particle transmission and beam spot size, for designing charged particle optics~\cite{huber2024cpopt}.
Researchers propose a fast multi-objective software framework using evolutionary algorithms to tackle the complex problem of designing and optimizing particle accelerators, such as the Argonne Wakefield Accelerator facility, to achieve optimal machine parameters and beam dynamics~\cite{PhysRevAccelBeams.22.054602}.
Researchers have developed a machine learning-based approach to create fast and accurate surrogate models for high-fidelity physics simulations of charged particle accelerators, enabling significant speedups in design studies and experiment planning~\cite{PhysRevAccelBeams.23.044601}.
However, GA is a purely optimization-focused algorithm as it leverages parallelized evolution, and does not allow continuous control for online accelerator operations. GA also requires re-running when there is a change in the linac, such as a cavity going offline. \\ \\
Bayesian Optimization has been explored for particle accelerator optimization~\cite{boltz2024more, hanuka_2021_physics, BO_review}.
Additionally, Multi-Objective Bayesian Optimization (MOBO), has been used online to efficiently map out Pareto-optimal tradeoffs between competing objectives~\cite{roussel_multiobjective_2021}. 

Unlike GAs, MOBO learns an underlying representation of the system behavior.
As long as the relevant inputs are being provided to MOBO, adjustments such as a cavity going offline can be accounted for with approaches such as contextual BO. 
While there are also avenues such as adaptive BO to account for changing conditions over time~\cite{kuklev2022online}, BO is primarily aimed at episodic optimization rather than continuous control, and becomes more computationally expensive as the number of sample points increases. 
A novel MOBO scheme is introduced to efficiently optimize particle accelerator performance online, reducing the number of observations needed to converge by at least an order of magnitude compared to current methods~\cite{PhysRevAccelBeams.24.062801}.
A MOBO approach has been used to optimize electron beam properties in the SLAC MeV-UED facility, allowing for faster and more efficient online beam tuning by searching the parameter space to identify the best trade-offs between key beam properties~\cite{Ji2024}\\ \\ 
In contrast, Reinforcement Learning (RL) is intrinsically a controls algorithm which can continuously learn and adapt with changing conditions while providing a real-time control~\cite{RL_control}.
There has been several studies using single objective RL for accelerator applications\cite{OnlineRL_BO, RL_control, RL_microbunching, RL_fewshot}, however, to the best of our knowledge there has not been any studies using Multi-Objective Reinforcement Learning (MORL) for particle accelerator optimization or controls.

\section{Problem Description}
\label{ch:problem}
CEBAF uses SRF cavities to accelerate electrons to a desired energy. 

When a RF field is applied, SRF cavities dissipate heat \(H_i\) into the cryogenic system according to Equation~\ref{eq:HeatLoad}. 
It is characterized by the cavity's current RF gradient \(G_i\), its length \(l_i\), impedance \(\omega_i\), and the quality factor \(Q_{i}(G_i)\) \cite{https://doi.org/10.48550/arxiv.1502.06877}.  
Given the non-linear nature of heat dissipation, poor choices in gradient distribution can result in considerably higher heat loads and reduced cryogenic system stability.
\begin{equation}
H_i = \frac{G_i^2   l_i}{\omega_i  Q_{i}(G_i)}    
\label{eq:HeatLoad}
\end{equation}

The length of cavities $l_i$, and their impedance $\omega_i$ are known and standardized to a high precision. 
The RF gradients $G_i$ are actively controlled during operation.
The quality factor $Q_i$ is a function of RF gradient.
It is only measured through time consuming procedures that would interrupt beam delivery.  
As such, quality factors for each cavity are determined only at cryo-module commissioning and through infrequent, ad hoc beam studies. 
Quality factor can be negatively impacted after commissioning through a variety of means, including the introduction of particulate contamination or frozen gases.  
How to parasitically (without beam intervention) determine these changes in quality factor is an open question complicated by the fact that the cavities in a cryogenic module share a single cryogenic helium supply and its associated control system.
As such, for this study we use a fixed value of $Q_i$ per cavity used by operations.

\begin{figure}
    \centering
    \includegraphics[width=0.6\linewidth]{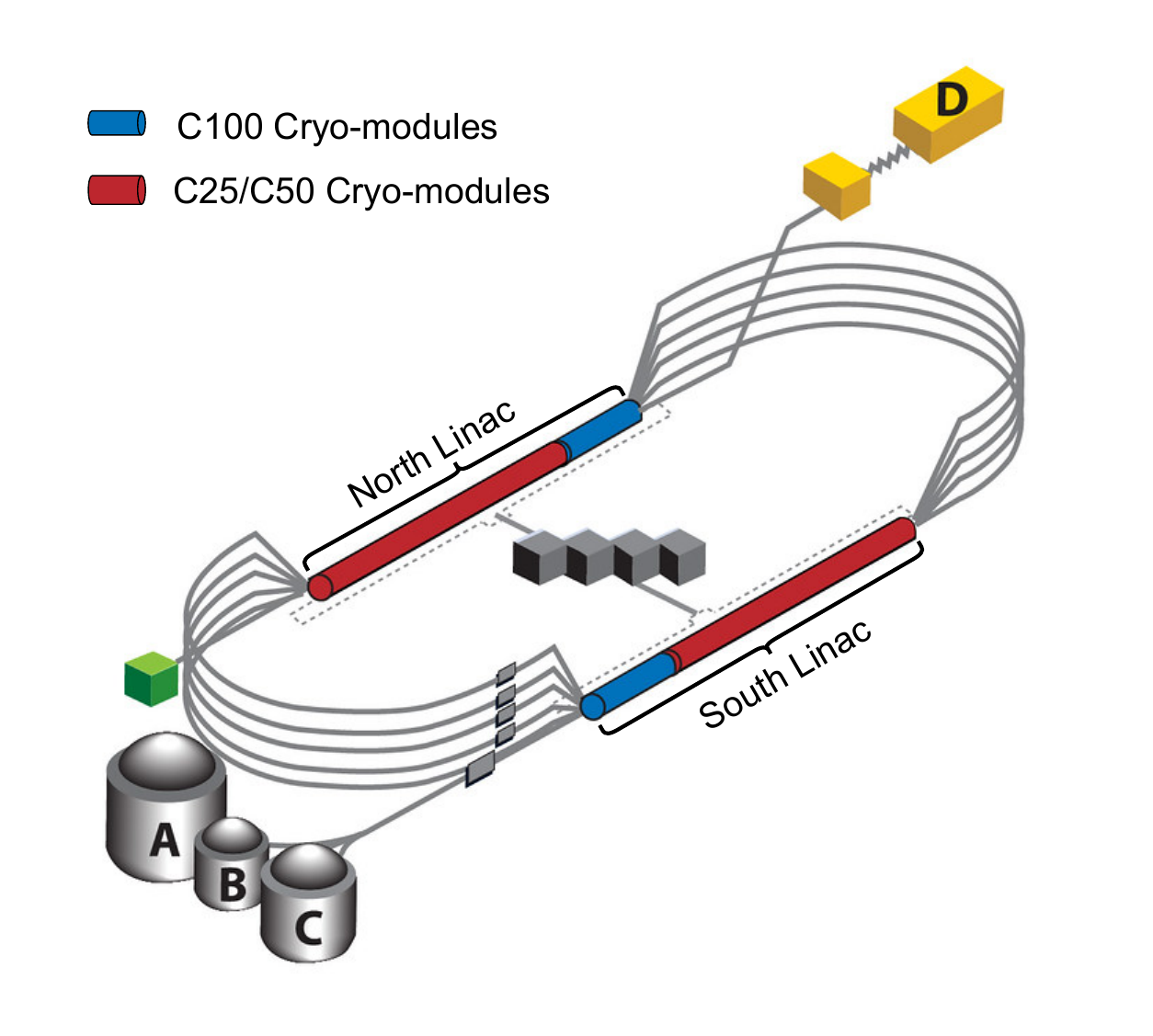}
    \caption{Schematic diagram of CEBAF with two anti-parallel North, and South Linacs to accelerate the electrons.}
    \label{fig:cebaf}
\end{figure}

Legacy cryogenic modules from CEBAF's 6 GeV era are still in operation and make up the majority of cryogenic modules in linacs.  
These are typically referred as C25 modules due to their median energy gain of 25 MeV per cryogenic module.  
One prevalent issue with C25s is that arc faults regularly occur as field emission from cavity walls charges a surface during RF operations and eventually discharges.  
When an arc fault is detected, the RF system for that cavity is turned off and the electron beam production is stopped to prevent potential damages.

Many C25s were refurbished into C50s in order to increase their median energy gain and mitigate this failure mode. The following family of log-linear statistical models, as described in Equation~\ref{eq:TripRates}, were developed using historical data that related a cavity's operating RF gradient to the arc fault trip rate, where \(T_i\) is the trip rate, \(G_i\) is the cavity RF gradient, \(F_i\) is the fault gradient, \(A\) and \(B_i\) are regression parameters.  \(F_i\) and \(B_i\) are fit from the data and \(A\) is set to -10.268 based on a previous study \cite{https://doi.org/10.48550/arxiv.1502.06877}.

\begin{equation}
T_i = e^{A + B_i(G_i - F_i)}
\label{eq:TripRates}
\end{equation}

In addition, the total energy gain (defined in Equation~\ref{eq:energygain}) across the linac needs to be a fixed value ($E_{linac}$) within a small tolerance ($\delta E$) in order to maintain proper steering and experimental requirements.
With the energy constraint, if one or more gradients are decreased, other cavities need to compensate by increasing their gradients. 
As each cavity has their own characteristic curves for heat load and trip rates, it introduces a trade-off between two objectives in an optimal solution set.
Finally, gradient distribution must account for unique operating characteristics of each cavity. For example, hardware failures may lock cavities to a specific gradient, and other limitations may restrict the range of RF gradients a cavity can sustain. 
These issues typically fall neatly into a distribution strategy as they are captured in a cavity's upper gradient limit.

\begin{equation}
    E = \sum_{i=0}^{N}l_i \times G_i
    \label{eq:energygain}
\end{equation}

From the above problem description, we formulate optimization problems using a physics based surrogate to model the RL heat load and FSD trip rate based on gradient distribution. 
The optimization problem is described in Equation~\ref{eq:opt}.
The optimization environment uses the OpenAI gymnasium framework~\cite{felten_toolkit_2023}, allowing seamless integration with different algorithms and packages.

\begin{eqnarray}
\label{eq:opt}
\mathrm{Minimize\ } &  &\ H(\boldsymbol{G}) = \sum_{i=1}^{N_{c}}H_{i}(G_i),\ T(\boldsymbol{G}) = \sum_{i=1}^{N_{c}} T_i(G_i)\nonumber \\
\mathrm{Subject\ to} &  &   \ |E_{linac}-\sum_{i=1}^{N_{c}}G_{i}l_{i}|<\delta E,\label{eq:opt_problem}\\
&  &\ \ a_{i}\leq G_{i}\leq b_{i}.\nonumber 
\end{eqnarray}

Here $H(G)$ and $T(G)$ are total RF heat load and FSD trip rate, $a_i$, $b_i$ are the lower and upper gradient limits respectively defined by the operational constraints, and $N_c$ is the number of cavities under consideration.

We formulated four optimization environments to gradually increase the problem complexity. 
The four environments are formed with 8 (single cryomodule), 16 (two cryomodules), 32 (four cryomodules), and 200 (entire north linac) RF cavities respectively for simultaneous optimization of heat load and trip rate.  
We used operational limits on individual cavity gradients as hard constraints.
The energy gain requirements are described in Table~\ref{tab:energy_constraint}.

\begin{table}[h!]
    \centering
    \begin{tabular}{|c|c|c|c|c|} 
    \hline 
       Env  & Number of Cavities & $E_{linac} \pm \delta E$ (MeV) & \multicolumn{2}{|c|}{Hypervolume} \\
       &&&Ref (H, T) & Ideal (H, T) \\
       \hline 
       \hline
       8D  & 8 & $20.08\pm0.40$ & (22.4, 0.05) & (20.9, 0.015)\\
       16D & 16  & $50.00\pm0.60$ & (100.0, 0.40) & (88.0, 0.015)\\
       32D & 32  & $120.00\pm0.80$ & (290.0, 0.40) & (262.0, 0.010)\\
       North linac & 200  & $1050.00\pm2.00$ & (2530.0, 6.00) & (2380.0, 1.000)\\
       \hline
    \end{tabular}    
    \caption{Optimization Environments; their state, and action dimensions, energy constraint, and reference points, and ideal points (lower bound) for hypervolume calculation}
    \label{tab:energy_constraint}
\end{table}

As the environment mainly consists of physics equations and is based on historical data, it can be implemented such that it is differentiable. We implement the environment in Tensorflow~\cite{tensorflow2015-whitepaper}, which allows auto-differentiation and backpropagation of gradients for DDRL training while other algorithms in this study do not leverage differentiability of the environment.

\section{Methods}
\label{ch:method}

\begin{figure}
    \centering
    \includegraphics[width=0.6\linewidth]{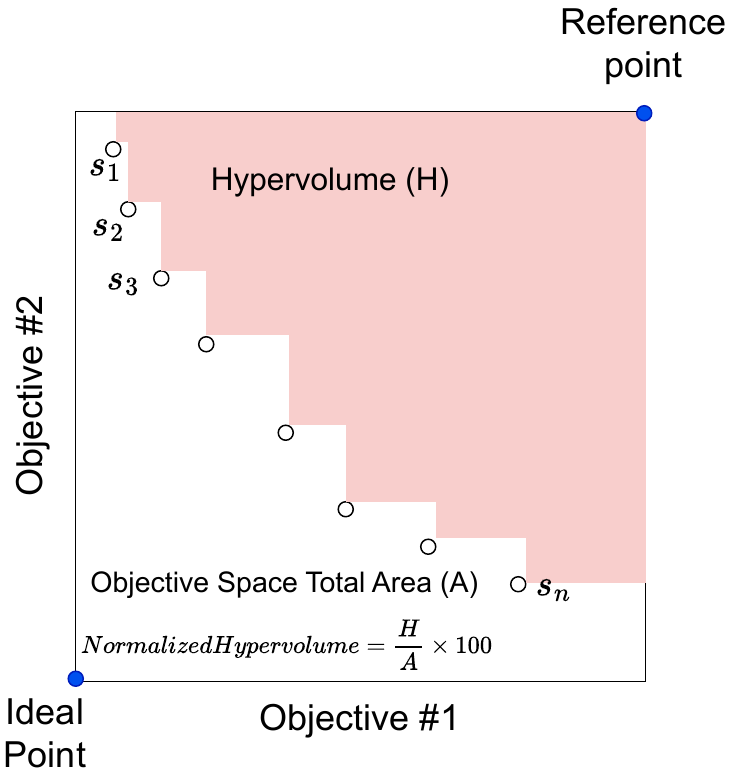}
    \caption{Hypervolume calculation in 2D and its normalization using ideal points}
    \label{fig:HVcalculation}
\end{figure}

Solutions to MOO problems are represented as a pareto front.
A parto-front represents a set of solutions where no objective can be improved without worsening another. In other words, it consists of all the optimal solutions that provide the best trade-offs among conflicting objectives.
Hypervolume (or S-metric)~\cite{Guerreiro_2021} is a measure used to evaluate the quality of a Pareto front. 
It quantifies the volume of the objective space that is dominated by a given Pareto front relative to a reference point.
The reference point is a point in the objective space that is typically worse than the best possible values for all objectives.
The hypervolume is calculated by measuring the volume of the region in the objective space that is dominated by the Pareto front and bounded by the reference point.
We normalize the hypervolume to percentage using the ideal points (lower bound of objective space). The reference and ideal points for each problem are listed in Table~\ref{tab:energy_constraint}.
We use hypervolume calculation provided by pymoo~\cite{pymoo} package to monitor the coverage during and at the end of evolution/training.
Figure~\ref{fig:HVcalculation} demonstrates the hypervolume calculation and normalization.
Using this hypervolume metric, we evaluate multi-objective optimization algorithms for our CEBAF problem. We describe each of these algorithms below.

\subsection{Genetic Algorithm }

GA is a type of optimization technique that mimics the process of natural evolution to find the best solution to a problem. 
It starts by creating a random initial population of potential solutions, known as individuals. Each individual is evaluated based on its fitness, or how well it solves the problem.
In each subsequent generation, the algorithm uses two processes to create new individuals: recombination and mutation. 
Recombination combines the characteristics of two existing individuals to create a new one, while mutation introduces random changes to an individual's characteristics. 
The new individuals are then evaluated and compared to the existing population.
The algorithm then selects the top half of the individuals with the best fitness values to move on to the next generation, effectively eliminating the less fit individuals. 
This process is repeated multiple times, allowing the fittest individuals to adapt and improve over time. 
As the algorithm iterates, it's likely to converge on a better solution, making it a powerful tool for solving complex optimization problems.

In this study, we employed the Non-Dominated Sorting Genetic Algorithm II (NSGA-II)~\cite{nsga2} to tackle a multi-objective optimization problem. NSGA-II is a fast and elitist genetic algorithm specifically designed for multi-objective problems, with a computational complexity of $O(MN^2)$ for a problem with $M$ objective functions and a population of size $N$.
When dealing with a population that involves multiple objective functions, individuals can be ranked and ordered using two key attributes: nondomination rank and crowding distance. An individual is considered non-inferior to another with the same nondomination rank, but inferior to one with a lower rank. When individuals have the same rank, they are ordered based on their crowding distance, with higher values indicating lower density in the solution space. This suggests that we should prioritize preserving individuals with higher crowding distances to maintain a diverse and well-spread population. NSGA-II can be briefly described as follows.  
\begin{enumerate}
	\item Create the first generation population randomly but ensure that each solution is within the constraints; sequence them by the nondomination rank. \\For a sequenced population of size $N$, we can generate the new population in the following two steps. 
	\item Create another $N$ individuals by recombination and mutation of the existing ones. 
	\item Sequence the $2N$ individuals by rank and discard the $N$ individuals with higher ranks. It is important to note that the constraints are folded into the fitness score. Any solution that violates the constraint(s) is assigned a low fitness score to not let them proceed to the next generation.\\

 Now we obtain the new generation population of size $N$.      
	\item Repeat steps 2 and 3 for a predetermined number. 
\end{enumerate}

\subsection{Bayesian Optimization}
Bayesian Optimization (BO) is a black-box method that traditionally has been used to optimize expensive, unknown functions~\cite{rasmussen_gaussian_2006}. 
More recently, BO has been used extensively in the context of online optimization of particle accelerators (see \cite{roussel_Bayesian_2024} for a summary). In BO, a probabilistic model is used to predict the mean and uncertainty of the function to be optimized based on the input variable values. 
This model is often composed of a Gaussian Process (GP), which is appealing because GPs are able to learn approximations of functions in a sample-efficient way (e.g. even from tens of samples) and inherently provide reasonable uncertainty estimates. 
During optimization, points are sampled and the GP is retrained at each iteration. An acquisition function is then used in conjunction with the GP model to choose a subsequent point to sample, which balances the uncertainty in the prediction with the likelihood of reaching a better optimum. 
One downside of GPs is that they do not scale as well as alternative models such as neural networks with respect to training data size, resulting in longer inference times that increase as the training dataset increases. 
This in turn affects the number of variables in a given optimization problem that can be solved in a time-efficient manner with GP-based BO. 
There are alternative approaches to improve scaling, such as using correlated and physics-informed kernels~\cite{duris_2020_Bayesian,hanuka_2021_physics}, Bayesian neural networks~\cite{Goan_2020} instead of GPs, and trust-region based BO \cite{trust-region-BO}; an overview can be found in \cite{roussel_Bayesian_2024}.

Here, we use constrained multi-objective BO (MOBO), which models each individual objective as independent GP and uses expected hypervolume improvement (EHVI)~\cite{BO_Book1} as the acquisition function. 

We also use learned output constraints, in which observed constraints are modeled as separate GPs and sampling is conducted with an acquisition function that balances the likelihood of objective improvement and constraint violation, according to the implementation described in \cite{daulton2021parallelbayesianoptimizationmultiple}. We use implementations in the Xopt package for MOBO \cite{Xopt} which is based on the implementation of MOBO in Botorch \cite{balandat2020botorch}.

One modification made to the MOBO algorithm in this case to handle the high dimensionality of the input space was to introduce a heuristic to optimize the acquisition function.
Generally, the acquisition function is numerically optimized from a set of random initial starting points.
This however can be a poor strategy in tightly constrained, high dimensional input spaces since the search space increases exponentially.
To address this challenge, we assume that the objective functions are relatively smooth, implying that the EHVI acquisition function will be maximized in a local region near previously observed Pareto-optimal points in input space.
We leverage this assumption by initializing acquisition function optimization at the location of Pareto-optimal points in input space if they are available (otherwise random points are chosen).
In empirical tests, this showed a significant reduction in the number of iterations needed to converge to the true Pareto front in the ZDT1 test problem \cite{10.1162/106365600568202}.
 
The returned energy, trip rate, and heat load from the optimization environment are used to assess whether constraints are violated, and the valid samples for trip rate and heat load from all previous interactions are used to calculate hypervolume of the Pareto front. The constraints are modeled separately from the objective, and both are taken into account when sampling new points.

\subsection{Reinforcement Learning}
Reinforcement Learning (RL) is a sub-field of Machine Learning.
The main components of RL are the agent and the environment. 
The agent is composed of the decision making logic which includes a policy.
The environment is the complex system that the agent interacts with.

RL agents repeatedly interact with the environment to learn how to make optimal decisions.
The environment returns a reward or penalty based on how favorable a given interaction was towards the optimal solution.
The environment can be an offline virtual physical, or surrogate model, or the real physical system.
In our case, the surrogate model captures the accelerator's behavior when changing the cavity gradients such as the heat load, trip rate, and beam energy.

RL is commonly used for control applications, however, it can be used for optimization for a static environment.

In a typical RL setting, the objective of the agent is to maximize the infinite sum of discounted future rewards. 
However, in this study, we are approaching the multi-objective problem as a purely optimization problem. 
Thus the agent is trained to optimize the environment in a single step limiting the exploration to short term (current) reward maximization.

Multi-objective reinforcement learning (MORL) research is pivotal in addressing the complex decision-making processes that require balancing multiple, often conflicting, objectives. 

For this problem, we defined the reward vector based on the two objectives in Equation~\ref{eq:reward}.
Here the reward vector $[R_h, R_t]$ is normalized between 0 and 1, and penalty P, is a relatively high value when the energy constraint is violated by a significant margin. 

\begin{equation}\label{eq:reward}
    R_h = -1 \times \sum_{i=0}^{N}H_i + P \quad and \quad R_t = -1 \times \sum_{i=0}^{N}T_i + P \\
\end{equation}

\begin{center}
  \sffamily
\begin{tabular}{l@{\,}l@{\quad}l}
  \ldelim\{{3}{*}[$P={}$]
 & $- 5 \times |E - E_{min}|$ & if $E < E_{min}$ \\
 & $- 5 \times |E - E_{max}|$ & if $E > E_{max}$ \\
 & 0 & Otherwise \\
\end{tabular}
\end{center}

Current RL approaches to solve multi-objective optimization problems are split between single-policy and multi-policy approaches~\cite{MORL_book}. 
We focus on a single-policy solution leveraging existing research, such as Preference Driven MORL (PD-MORL)~\cite{PD-MORL}.
For this paper, we study two implementations of RL described in Sections~\ref{sec:cmorl} and~\ref{sec:cddrl}.

\subsubsection{Conditional Multi-Objective Twin-Delayed Deep Deterministic Policy Gradient (CMO-TD3)}
\label{sec:cmorl}
\hfill\\
The first implementation is based the TD3~\cite{TD3}, with two key differences:
\begin{enumerate}
    \item The policy, $\pi_{\theta}(s|\alpha)$, takes a conditional probability vector ($\alpha$) defining the priority for each objective
    \item The critic predicts a q-value vector providing the value for each objective separately
\end{enumerate}
As shown in Algorithm~\ref{alg:grl_algo}, the policy optimization collapses into a single objective when taking the dot product of the conditional and critic's $Q$-value vectors.
 
An example of this would be a linear combination of the rewards with the weights of each objective being proportional to the importance of the objective to the end user. 
This approach finds a single point on the Pareto front as the agent is optimizing for a single static combination of the objectives. 
This approach allows the policy to learn the full Pareto front by changing the conditional probability input $\alpha$.

\begin{algorithm}
\caption{Conditional Multi-Objective TD3}
\label{alg:grl_algo}
\SetAlgoLined
Initialize critic networks $Q_{\theta_{1}}$, $Q_{\theta_{2}}$ and policy network $\pi_{\phi}$\\
Initialize target networks $\theta_{i}'\leftarrow \theta_{i}$ and $\phi'\leftarrow \phi$\\
Initialize the replay buffer $\mathcal{B}$\\

\For{ t in 1,...,T, }{
    Select action $a$ using the policy $\pi_{\phi}(s|\alpha)+\epsilon$; $n\sim\mathcal{\epsilon(0,\sigma)}$; $s$ is the current state; and $\alpha$ is the conditional parameter that defines the priority of each objective.\\
    Observe the next state ($s'$) and associated rewards ($r$)\\
    Store transition tuple $(s, a, r, s', \alpha)$ in $\mathcal{B}$\\ 
    Randomly sample N transitions from $\mathcal{B}$\\
    $a' \leftarrow \pi_{\phi'}(s'| \alpha) + n;$ \\
    $y_{i} \leftarrow r + \gamma$min$_{i} Q_{\theta_{i}'}(s',a')$\\
    Update critics $\theta_{i} \leftarrow N^{-1} \sum (y_{i}-Q_{\theta_{i}}(s,a))^2$\\
    \If{ t mod d}{
        Update conditional policy gradient:\\
        $\nabla_{\phi}J(\phi) =  N^{-1} \sum \nabla_{a} [\alpha \cdot Q_{\theta_{1}}(s,a)|_{a=\pi_{\phi'}(s|\alpha)}]$\\
        Update target networks:\\
        $\theta_{i}'\leftarrow \tau \theta_{i}' + (1-\tau)\theta_{i}$\\
        $\phi'\leftarrow \tau\phi' + (1-\tau) \phi$
        }
    }
\end{algorithm}

\subsubsection{Conditional Multi-Objective Deep Differentiable Reinforcement Learning (CMO-DDRL)}\label{sec:cddrl}
\hfill \\
\begin{algorithm}[h]
\caption{CMO-DDRL}
\label{alg:ddrl_algo}
\SetAlgoLined
Initialize actor policy $\pi_{\phi}$\\
Initialize the policy learning rate $l = L_{0}$\\
\For{ t in 1,...,T, }{
    Select action $a$ using the policy $\pi_{\phi}(s|\alpha)$;
    \textit{where $s$ is the current state; and $\alpha$ defines the priority of each objective.} \\
    Observe the next state ($s'$) and associated reward vector ($r$)\\
    \textbf{Training Loop;} \\
    Sample $\alpha \sim Dirichlet({\alpha_1, \dots, \alpha_k})$; where k=number of objectives\\
        $a \leftarrow \pi_{\phi}(s| \alpha)$ \\
        $s', r \leftarrow env(a)$ \\
        Update conditional policy gradient:\\
        $\nabla_{\phi}J(\phi) =  N^{-1} \sum \nabla_{a} [\alpha \cdot r(a)|_{a=\pi_{\phi'}(s|\alpha)}]$\\
    \If{ $t mod d = 0 and l > l_{min}$ }
    {
    Anneal learning rate $l \leftarrow l/\beta$;
    \textit{where $\beta$ is learning rate reduction factor}
    }
}
\end{algorithm}
Differentiable techniques are very powerful and sample efficient in learning the concepts~\cite{huge2020differentialmachinelearning} as compared to traditional RL algorithms that mainly rely on exploration by sampling the environment.
In contrast to the model-free CMO-TD3 algorithm, DDRL relies on the environment being differentiable and back-propagates the short (or long) term reward through the environment to update the agent weights and biases.

The agent interacts with the environment analogous to TD3, however it does not require critic model or a buffer for storing experiences as it can utilize the reward from the environment directly for backpropagation. 
The reward can be formulated as either short term or long term. 
As described in Algorithm~\ref{alg:ddrl_algo}, the agent observes the current state of the environment, generates a batch of actions conditioned on conditional probability input $\alpha$, applies them to the environment to compute the reward (single-step) and next step. 
Subsequently, the agent employs the reward and utilizes automatic differentiation to update its weights and biases to optimize reward maximization. 
It is important to note that this agent is used for one-step optimization, however, it can be used for sequential control problems modeled as a Markov Decision Process (MDP) by unrolling $N$ number of future steps as described in a recent study~\cite{DDRL}.

\section{Results}
\label{ch:results}

We evaluate MOGA, MOBO, CMO-TD3, and CMO-DDRL algorithms on the MOO environments described in Section~\ref{ch:problem} and compare their performance in terms of quality of solutions and convergence time.
On all the problem environments, we first benchmark the results using MOGA algorithm as described in appendix A.
To provide a statistically robust analysis, we conducted 16 trials for each algorithm using unique random seeds.
Each trial was initiated using a unique random seed.
We present the summarized results by computing the $2\sigma$ confidence bounds using the corresponding quantiles of 0.023 and 0.977 for the lower and upper bounds, respectively.

\begin{figure}
    \centering
    \includegraphics[width=0.8\linewidth]{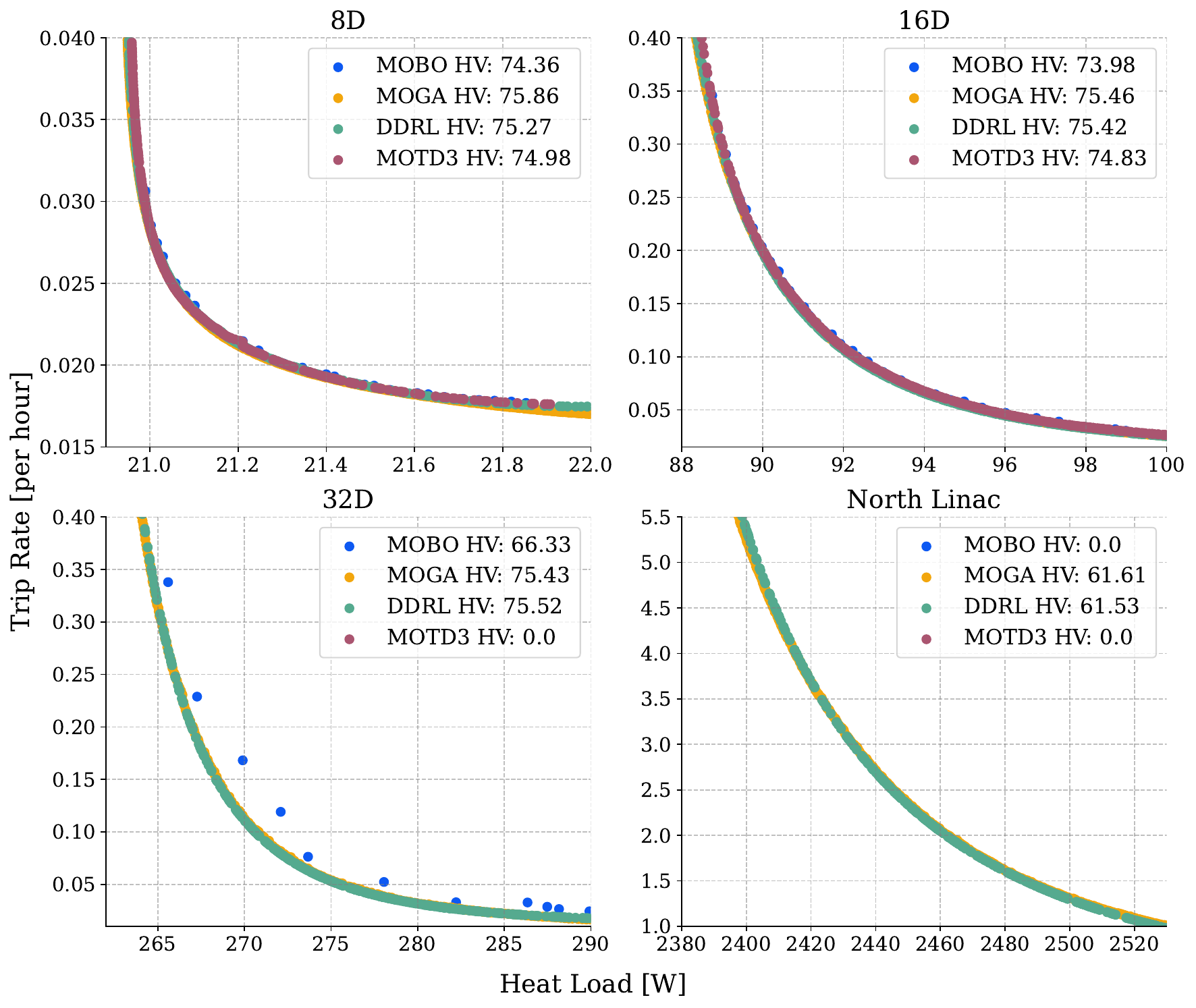}
    \caption{Comparison of optimal Pareto front produced by MOBO, MOGA, MOTD3 and DDRL algorithms. 
    The best Pareto fronts are chosen for each algorithm from 16 trials. Hypervolumes are displayed in the legend.}
    \label{fig:paretoComparisonAll}
\end{figure}

\begin{figure}[h]
    \centering
    \includegraphics[width=0.5\linewidth]{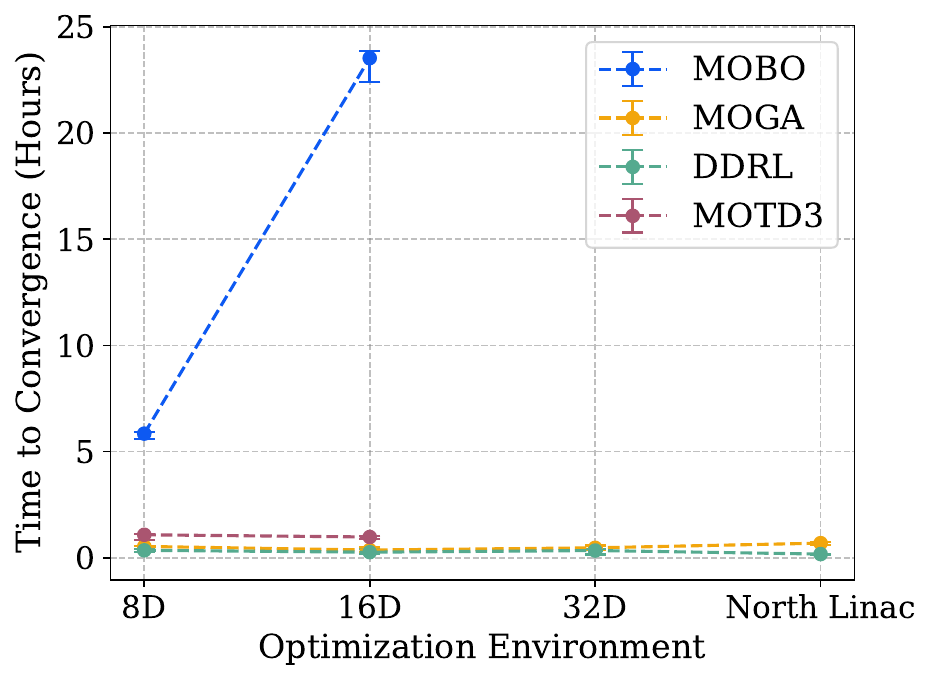}
    \caption{Time taken to converge; Each scatter dot represent median and the error bars cover 2$\sigma$  confidence bound over 16 trials.}
    \label{fig:convergenceTime}
\end{figure}

\begin{figure}[h]
    \centering
    \includegraphics[width=0.8\linewidth]{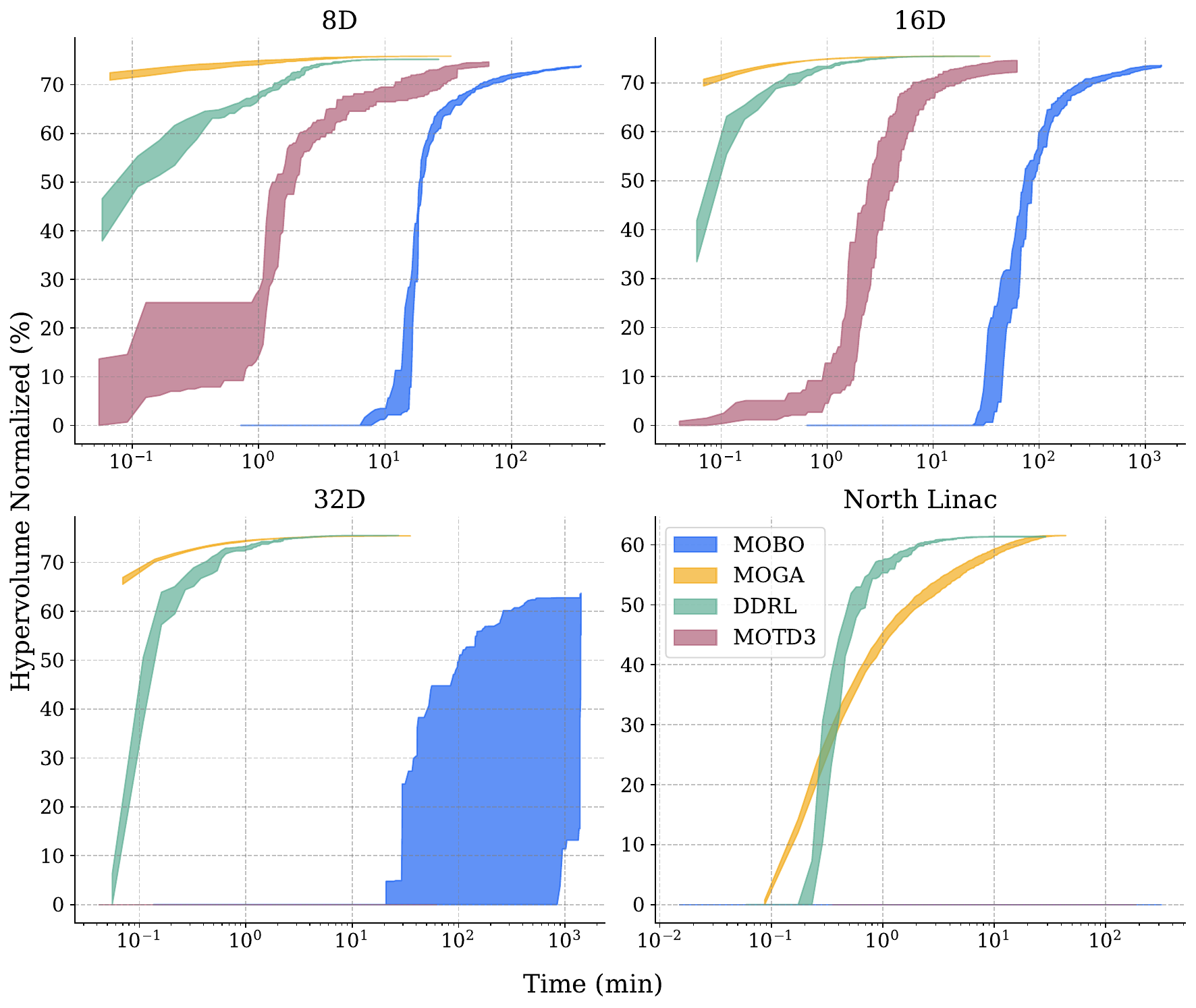}
    \caption{Pareto Coverage over time on different environments with the four algorithms. x-axis (log scale) represents time in minutes. While MOBO takes longest to converge, DDRL is the fastest on large scale problem (200-dimensional). See also the corresponding convergence plot in terms of samples/iterations in Figure~\ref{fig:ParetoCoverageStep}. Note - The time axis is shown in log scale and does not include hypervolume at 0 seconds which is zero for all the algorithms due to random initialization.}
    \label{fig:ParetoCoverageTime}
\end{figure}

\begin{figure}
    \centering
    \includegraphics[width=0.8\linewidth]{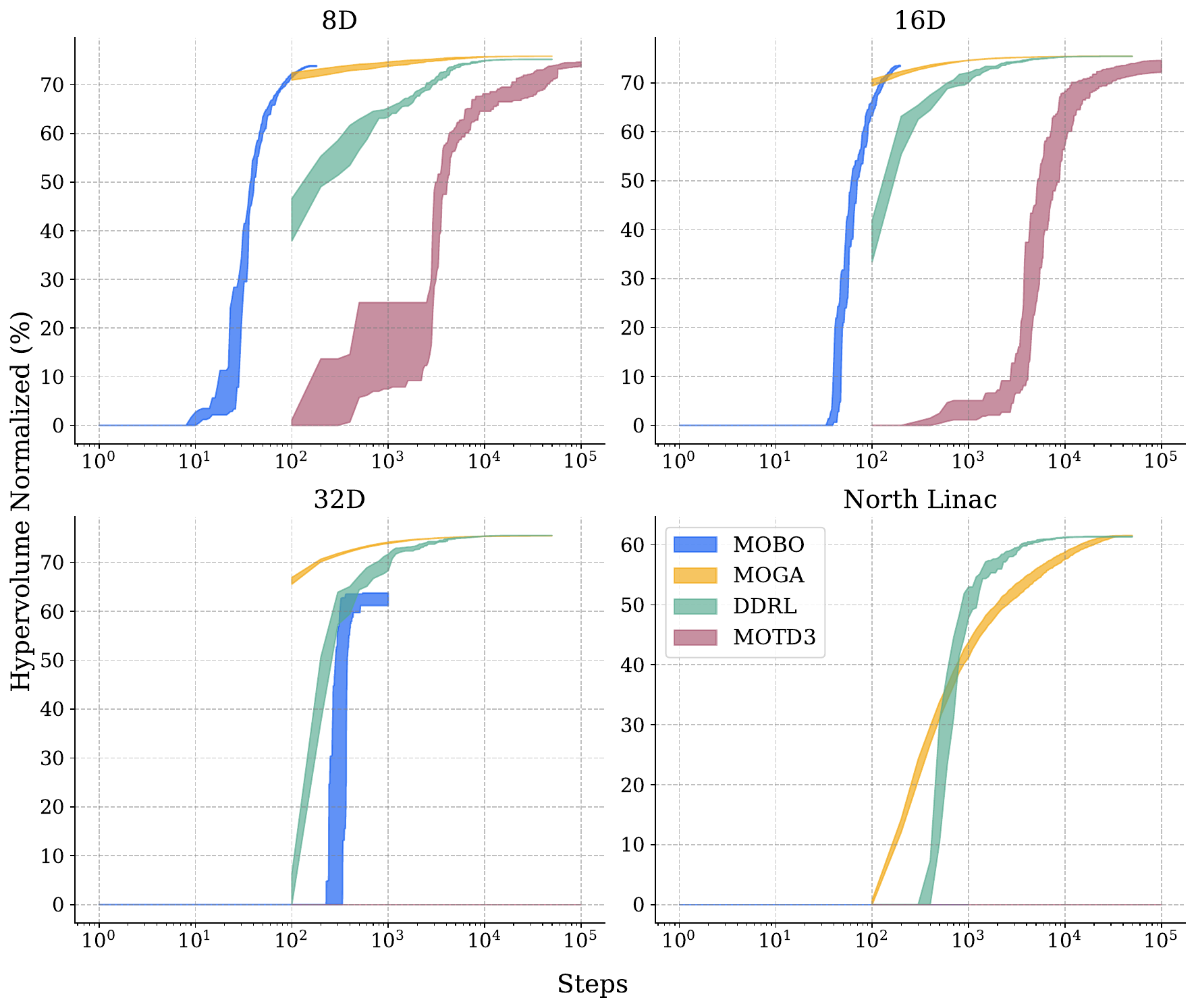}
    \caption{Pareto Coverage on different environments with the four algorithms. Note -  the x-axis is shown in log scale and does not include iteration 0, as such the Hypervolume at that step (zero due to random initialization) is not shown in the figures.}
    \label{fig:ParetoCoverageStep}
\end{figure}

On the smaller scale problems (8, and 16 dimensional), all four algorithms converge to a similar Pareto front.
However, as the size of the state and action space is scaled up to 32 and 200 (North Linac), CMO-DDRL and MOGA converge to similar optimal solutions while none of the trials for MOBO and CMO-TD3 could converge in a given time span.
Figure~\ref{fig:paretoComparisonAll} shows there is no Pareto front for CMO-TD3 and MOBO for north-linac optimization as they produce infeasible solutions that are outside the energy constraint and do not converge.
To demonstrate the evolution of these algorithms during training, Figure~\ref{fig:ParetoCoverageTime} shows hypervolume coverage with time, and Figure~\ref{fig:ParetoCoverageStep} shows the hypervolume coverage per algorithm step. This highlights the differences between the algorithms in terms of both time-efficiency and sample-efficiency.

For MOGA, CMO-DDRL, and CMO-TD3, we record hypervolume at every 100 steps whereas for MOBO algorithm we record it at every step, due to the smaller number of overall steps taken and greater computation time per step relative to the other algorithms.

It is important to note that the convergence time shown in Figure~\ref{fig:ParetoCoverageTime} and Figure~\ref{fig:convergenceTime} only account for algorithm training/evolution time and do not include hypervolume calculation (if not required by the training step) or results saving.
In terms of time-efficiency, it is evident that MOGA is faster on our smaller scale problems (16 and 32 -dimensional) than all the other algorithms, where as on the  32D problem the convergence speed of MOGA is comparable to CMO-DDRL, and on the larger scale problem of 200D, CMO-DDRL is fastest to converge. 

MOBO is slowest to converge on all the optimization environments.
This is also evident in Figure~\ref{fig:convergenceTime} that displays median convergence time taken by each algorithm with 2$\sigma$ confidence bounds.

On the other hand, MOBO takes significantly fewer iterations to converge on the smaller scale problems, as can be seen in Figure~\ref{fig:ParetoCoverageStep}. It also scales poorly with the state and action space size. 
As we go from 8-dimensional to 16-dimensional problem, the time to convergence increases to almost four times. 
This highlights that in cases where the time cost of acquisitions is high and problem dimensionality is modest, the sample-efficiency of MOBO makes it appealing; however, in cases such as ours where the time cost of acquisitions is very low, CMO-DDRL and MOGA provide overall greater time-efficiency despite their lower sample-efficiency.  
This is an important consideration which algorithm to use for other problems. 

Notably, even though MOGA converges, it does take many generations to converge compared to some other problems in accelerators (e.g. see \cite{Edelen_2020_machine}), especially for the 200D North Linac version of the problem. Observing the MOGA convergence, starting from random sampling the algorithm moves slowly from an initial infeasible region (due to the energy constraint) and eventually reaches the valid region of parameter space. We also note that extensive random sampling of the parameter space for the 200D case does not produce valid samples. This leads us to speculate that one reason some algorithms, such as MOBO and MOTD3, may struggle to converge in this case is that the optimum is a very narrow one in a very large parameter space. MOGA through many parallel samples eventually converges, and DDRL can take advantage of the differentiable model to aid the direction of search. While techniques such as trust region BO can aid optimization of deep, narrow minima \cite{roussel_Bayesian_2024}, in this case we speculate the high dimension of the parameter space, coupled with starting very far in parameter space from the minimum, make it challenging to converge. Techniques that could potentially improve performance include  using priors or a correlated kernel from the differentiable model \cite{}, starting closer to the valid region of parameter space (as is often possible when running online), or other approaches to reduce the computational cost of the GP modeling or acquisition function optimization. We plan to study this further in future work.

One of the key challenges for the current implementation of MORL is intrinsic sample inefficiency for the default OpenAI gymnasium sampler when applied to a high dimensional global constraint problem. 
This problem is exacerbated when considering the penalty in the reward structure since the probability of a valid action is effectively zero and forces the agent to optimize in the phase space outside the energy constraint.

From these results, DDRL preforms better for large dimensional control applications because the algorithm can quickly converge to an optimal solution by leveraging back-propagation. 

In contrast MOTD3 is relatively challenging to train due to its reliance on exploration and sample inefficiency. 
On the other side, MOGA, and MOBO are suitable for optimization problems. 
However, in terms of time-efficiency, MOBO is relatively slow in this setting where the environment execution is very fast. As the higher sample-efficiency shows, MOBO is more suitable in environments where the sampling time (or execution speed of the environment) is higher.

In considering how these algorithms would translate to use on the actual accelerator, as opposed to the offline optimization problem described here, it is important to consider the rate at which settings and read-backs can be made and how close the surrogate model behavior is to the real machine behavior. One reason BO is popular for online optimization is because it can be used in cases where there is no previous data available; if starting from a good operating point on the accelerator, it may not struggle to converge in the same way observed in this broader offline optimization problem. BO also directly takes into account noise and other sources of uncertainty that occur on real-world systems. For DDRL and MOTD3, if the surrogate system is close enough to the real accelerator behavior, the algorithms may be able to be trained offline, as was done here, and transferred to the real system without substantial additional retraining. This prospect would require further study to assess, but the fact that the initial surrogate model used in training is partially data-driven lowers some of the risk in this process. The ability to retrain the RL agent depends on the sampling speed and data availability. On the real system, heat load can be quickly inferred from the cryogenic system (though it may be noisy); in contrast, trip rates are only evaluated theoretically based on historical data and would need to be approximated.

Finally, another major difference between the algorithms is that MOGA uses parallel execution in sample acquisition, which is not possible in a real-world system. In contrast, DDRL, TD3 and MOBO however use serial acquisition, meaning both can also be used in the real-world optimization/control setting. 

\subsection{Conditional Tuning for Multi-Objective Solutions}
\begin{figure}
    \centering
    \includegraphics[width=0.8\linewidth]{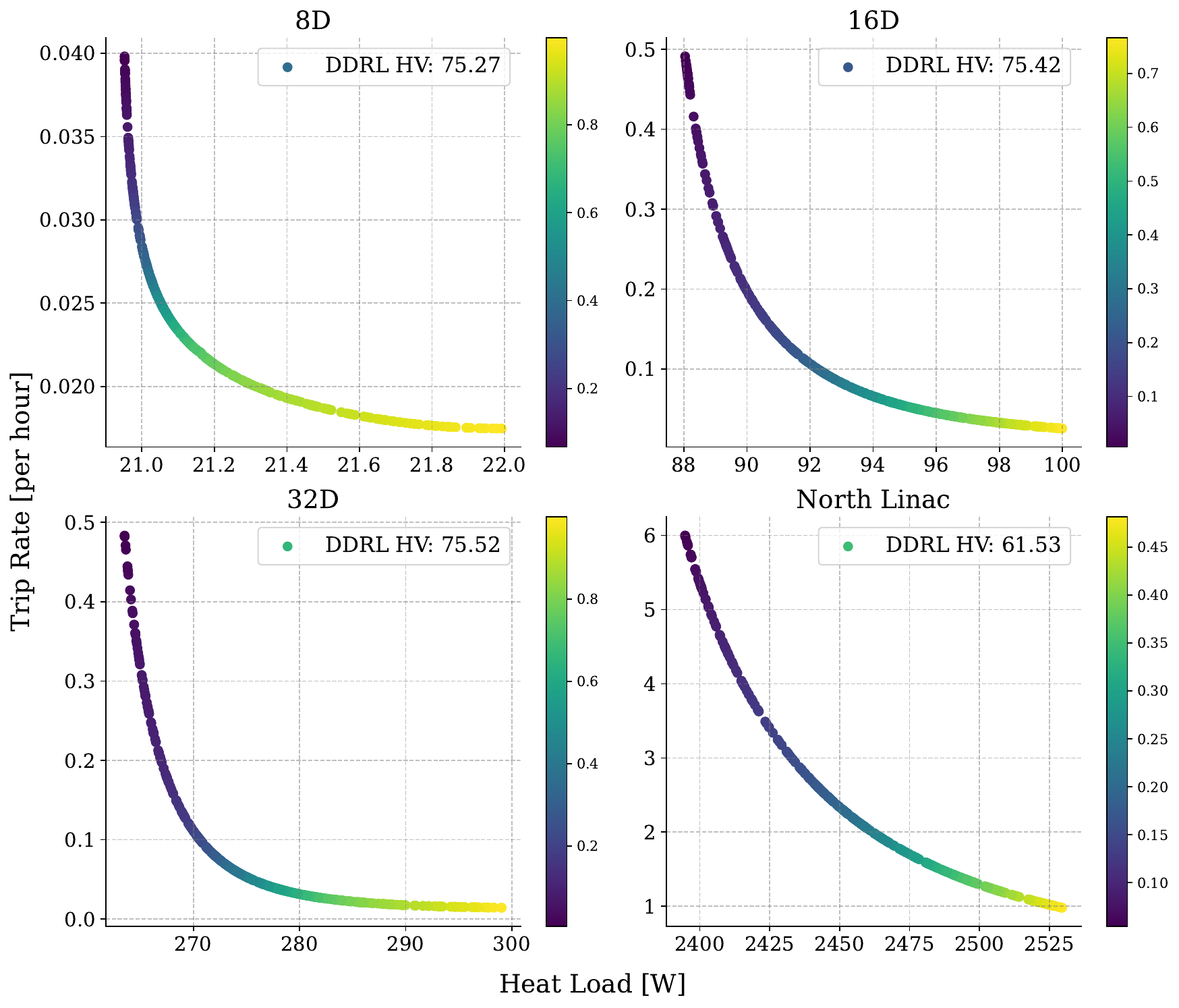}
    \caption{Tunable conditional parameter $\alpha$ for Pareto-optimal solutions from DDRL}
    \label{fig:ddrl-alpha}
\end{figure}

\begin{figure}
    \centering
    \includegraphics[width=0.8\linewidth]{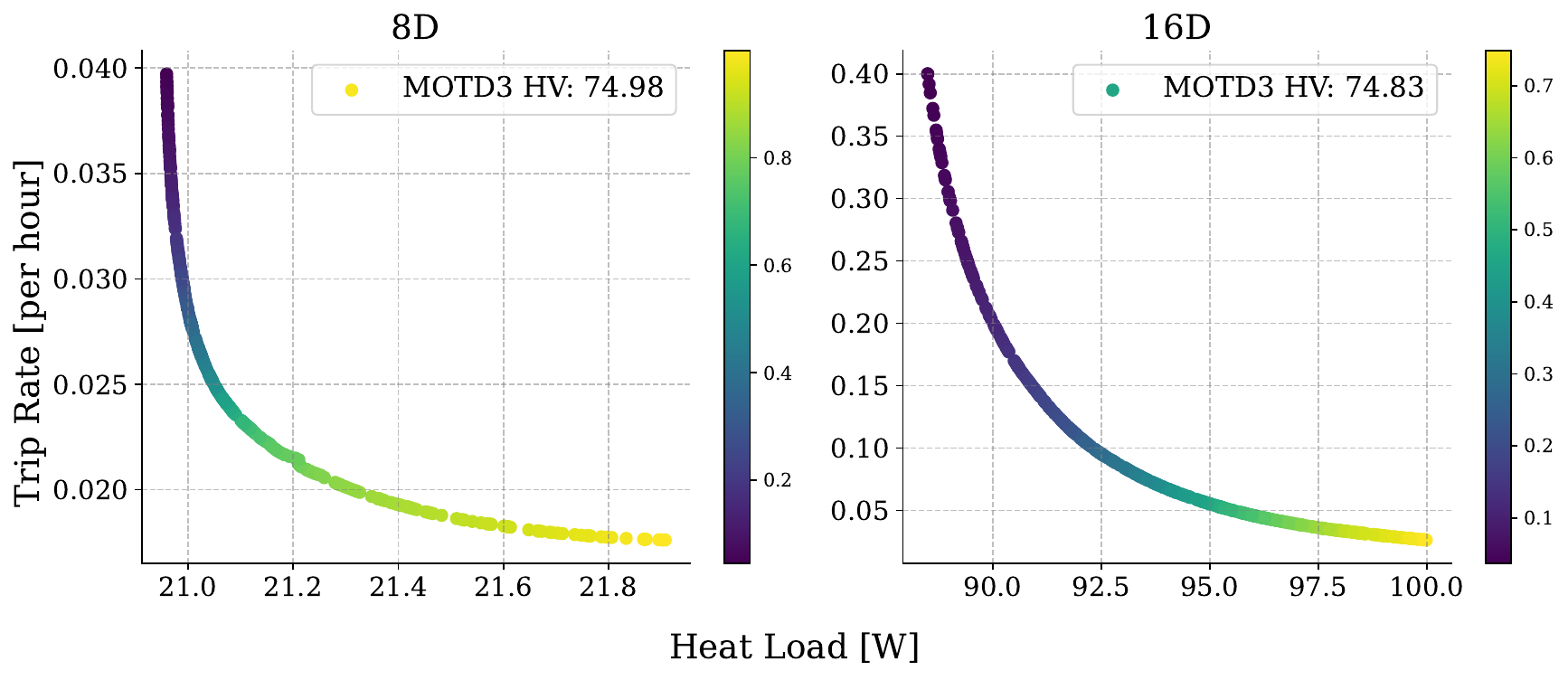}
    \caption{Tunable conditional parameter $\alpha$ for Pareto-optimal solutions from MOTD3}
    \label{fig:motd3-alpha}
\end{figure}

Optimal solutions to multi-objective optimization problems are produced in the form of a Pareto-front. 
It is often required to prioritize one or more objectives over others.
This is particularly important for deployed policies that can provide operators a tunable knob to focus on different objectives as demanded by operational needs.
Both MOTD3 and DDRL policies use a conditional input to produce different solutions on the optimal pareto-front.
The solutions produced by conditional policies are correlated with the optimal solutions on Pareto-front.
Figure~\ref{fig:ddrl-alpha}, show the Pareto fronts produced by the DDRL algorithm with the colorbar representing conditional input.
It is clear that the distribution of the optimal solutions follow a monotonic behavior with $\alpha$, making $\alpha$ a tunable knob to get to different solutions on the optimal Pareto-front.
This behavior is consistent with MOTD3, as shown in Figure~\ref{fig:motd3-alpha} for the two environments where it converges.

\section{Conclusion}
\label{ch:conclusion}
In this paper, we compared four algorithms (CMO-DDRL, CMO-TD3, MOBO, and MOGA) on a MOO problem to simultaneously optimize RF heat load and FSD trip rate via gradient distribution in CEBAF linacs constrained by an energy requirement.
Four optimization problems are formulated to show how this comparison depends on problem complexity, which scales from a single cryomodule to the entire CEBAF North linac.
We demonstrated that CMO-DDRL leverages differentiability of the optimization environment to outperform other algorithms in convergence speed on high dimensional problem. On smaller scale problems, all four algorithms reach similar solutions. 
However, convergence speed comparison reveals that MOBO is slowest to reach optimal solution even though being sample efficient.
It is important to note that MOBO and CMO-TD3 did not converge on the larger scale CEBAF problems, we suspect due to the high dimensionality of and particular difficulty of the global multi-dimensional hard constraints in this case.
Potential future work can include enhancing CMO-TD3 warmup on large scale problems with better random sampling that can include hard constraints, and optimizing MOBO in various ways to deal with the high dimensional constrained problem.
Another avenue of interesting future research is usage of MORL algorithms for sequential control, especially DDRL and evaluation of its capabilities in handling dynamic environments.

\section{Acknowledgement}
\label{ch:ack}
This research was supported by the U.S. Department of Energy, through the Office of Advanced Scientific Computing Research’s “Data-Driven Decision Control for Complex Systems (DnC2S)” project. Contributions from A.E. and R.R. were supported by the Department of Energy, Laboratory Directed Research and Development program at SLAC National Accelerator Laboratory, under contract DE-AC02-76SF00515.
This manuscript has been authored by Jefferson Science Associates (JSA) operating the Thomas Jefferson National Accelerator Facility for the U.S. Department of Energy under Contract No. DE-AC05-06OR23177. 
The authors acknowledge support by the U.S. Department of Energy, Office of Science. 
The US government retains and the publisher, by accepting the article for publication, acknowledges that the US government retains a nonexclusive, paid-up, irrevocable, worldwide license to publish or reproduce the published form of this manuscript, or allow others to do so, for US government purposes. DOE will provide public access to these results of federally sponsored research in accordance with the DOE Public Access Plan (http://energy.gov/downloads/doe-public-access-plan)

\section*{References}
\bibliographystyle{iopart-num}
\bibliography{references, moo_references}

\newpage
\appendix
\label{appendix:nsga_benchmark}

\section{Benchmark with NSGA-II}

\begin{figure}[h]
    \centering
    \includegraphics[width=0.8\linewidth]{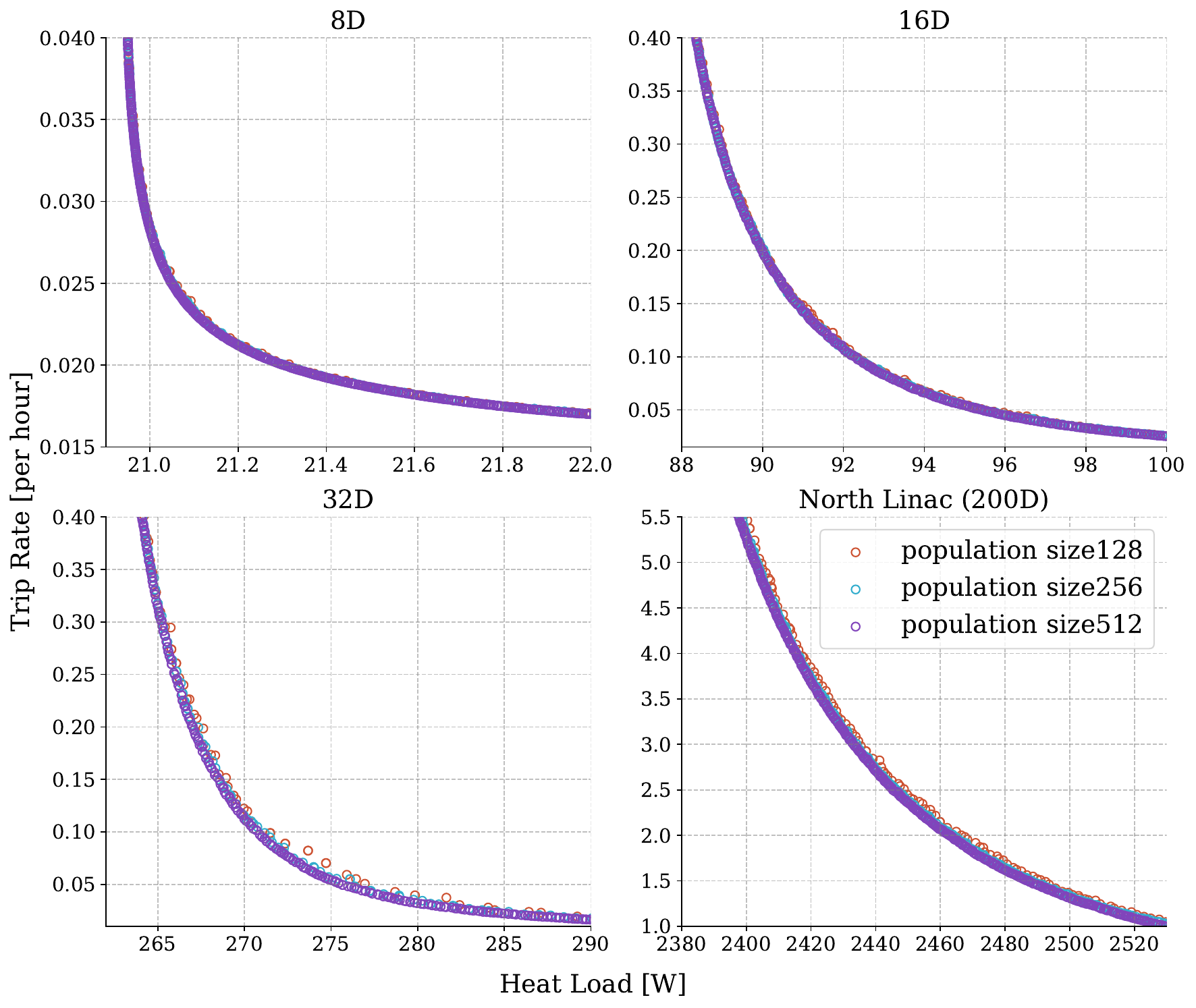}
    \caption{Comparison of Pareto-front obtained from NSGA-II with varying population sizes at 50K generation}
    \label{fig:GA_Comparison}
\end{figure}

\begin{figure}
    \centering
    \includegraphics[width=0.8\linewidth]{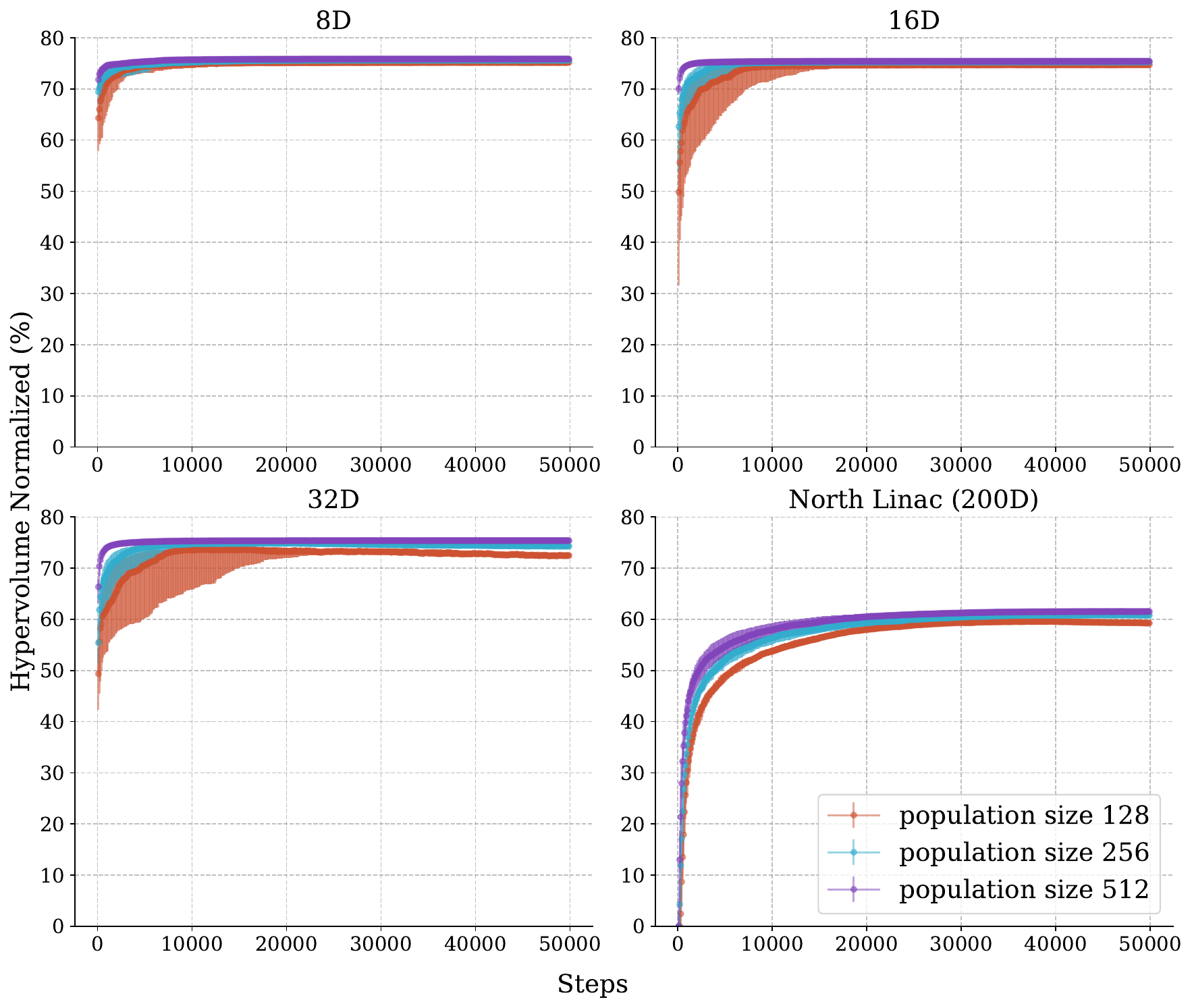}
    \caption{Hypervolume coverage of the pareto-front produced by NSGA-II during evolution. The area plots represents $2-\sigma$ confidence bound around median over 16 trials with random initial seeds.}
    \label{fig:GA_Coverage}
\end{figure}

In order to benchmark the pareto-front, we run the NSGA-II algorithm with varying population size and monitor the pareto-front coverage with the hypervolume.
For simplicity, we fix the mutation and crossover probabilities to 0.01 and 0.95 chosen after a light hyper-parameter tuning.
We run NSGA-II with population sizes of 128, 256, and 512 and observe that 
all the three variation of NSGA-II algorithm converges to similar solutions with small differences.
In general, running with larger population size generally produce higher hypervolume for the same number of iterations.
This is evident in figure~\ref{fig:GA_Coverage}, the figure shows hyper-volume coverage for NSGA-II evolution.
We limit the number of generations to 50K as the hyper-volume coverage plateaus.
The hypervolume is slightly higher with increasing number of population size as each point contribute to the volume (area in this case). 
However, as seen in Figure~\ref{fig:GA_Comparison}, the optimal pareto-front solutions with the three population sizes at 50K generation is similar for smaller scale (8 and 16 -dimensional) environments, however, as we scale up the action and state space (32, and 200 -dimensional problems), a larger population size of 512 produce better optimal solutions. 
As such, we use NSGA-II with population size of 512 for comparison with other algorithms.

\end{document}